\documentclass[
    ,final            
  ]
  {aipproc}

\usepackage{amsmath}
\usepackage{amssymb}
\usepackage{graphicx}
\usepackage{color}
\usepackage{ulem}

\layoutstyle{6x9}

\begin{document}

\title{TMD-factorization, Factorization Breaking and Evolution}


\author{S.~M.~Aybat}{
  address={Department of Physics and Astronomy,\\ Vrije Universiteit Amsterdam,\\ NL-1081 HV Amsterdam, The Netherlands}
,altaddress={Nikhef Theory Group, \\ Science Park 105, 1098XG Amsterdam, The Netherlands} 
}

\author{T.~C.~Rogers}{
  address={Department of Physics and Astronomy,\\ Vrije Universiteit Amsterdam,\\ NL-1081 HV Amsterdam, The Netherlands}
}

\begin{abstract}
We give an overview of the current status of 
perturbative QCD factorization theorems in processes that 
involve transverse momentum dependent (TMD) parton distribution functions (PDFs)
and fragmentation functions (FF).  
We enumerate those cases where TMD-factorization is well-established, 
and mention cases where it is likely to fail.  
We discuss recent progress in the implementation of specific TMD-factorization calculations, including the implementation of evolution.  We also give examples of hard part calculations.  
We end by discussing future strategies for the implementation 
of TMD-factorization in phenomenological applications. 
\end{abstract}

\maketitle
\vspace{-11.5cm}
\hspace{107mm}
NIKHEF-2011-022
\vspace{11.5cm}


\section{Complications with TMD-Factorization}
Many of the complications characteristic of TMD-factorization 
derivations are related to the difficulty
of establishing unambiguous definitions for the TMD PDFs and FFs (TMDs).  
The most natural attempts at definitions 
lead to divergences like the well-known 
"rapidity divergences" and Wilson line self-energies~\cite{Collins:2008ht}.  Moreover, a complete 
TMD-factorization derivation must account for soft gluons (gluons with nearly zero center-of-mass rapidity)
which give rise to soft factors in the factorization formula.  

Confusion over definitions is often manifested in questions about the appropriate
Wilson lines or "gauge links" that are needed to make TMDs gauge invariant.  
In many cases, the general structure of the Wilson line can be anticipated from very loose 
considerations of the process.  Formally, however, the structure of the Wilson 
must follow from a factorization derivation.  
A central theme of our talk is that the precise TMD definitions the should be used for phenomenology are ultimately 
dictated by the requirements of factorization.

A comparison between semi-inclusive deep inelastic scattering (SIDIS) and the Drell-Yan (DY) process 
illustrates the phenomenological importance of the Wilson line.  
TMD-factorization is valid in both cases, but 
in SIDIS the factorization derivation leads to a future-pointing Wilson line in the definition of the TMD PDF, whereas in DY it is past-pointing.  The result is a sign-flip acquired by T-odd functions in DY as 
compared to SIDIS~\cite{Collins:2002kn}.

In the classic electroweak processes ($e^+ + e^- \to H_1 + H_2 + X$, SIDIS, and DY), TMD-factorization 
derivations exist, and the TMDs are universal, up to the possible minus signs for T-odd functions.  
However, in processes that involve more than one hadron in the initial state, with observed hadrons in the 
final state, there are interesting 
effects that make the treatment of Wilson lines much more delicate. 
In factorization proofs, contour deformations
must be applied to momentum 
integrals before the approximations leading to a factorization formula are justified.  
Afterward,
Ward identities allow soft and collinear 
gluons to be pulled into factors that correspond to the Wilson line operators in the 
definitions of the TMDs~\cite{collins}.  However, in processes like
$H_1 + H_2 \to H_3 + H_4 + X$
the deformations must be made in different directions for different graphs
depending on whether they correspond to initial or final state interactions.  
Therefore, the usual Ward identity arguments do not apply, even 
though eikonal factors reminiscent of Wilson line contributions
do arise graph-by-graph.
At a minimum, a more complicated Wilson line 
structure is needed for the definitions of the TMDs in order to have factorization.
Recalling the non-universal sign of T-odd functions 
in the comparison between SIDIS and DY, one might hope to still make 
predictions by studying how non-universal Wilson line structures affect the TMDs. 
However, a detailed consideration of TMD-factorization for 
these processes 
reveals that TMD-factorization fails even if the TMD definitions are allowed to contain 
non-universal Wilson line structures~\cite{Rogers:2010dm}.  The problems arise because the gluons radiated from 
one of the incoming hadrons, of a type that contribute to a Wilson line, are affected via non-trivial color flow by 
gluons radiated from the \textit{other} hadrons.  (This is sometimes called "color entanglement.")

To summarize the above, we tabulate the status of TMD-factorization for the 
processes considered above, with a "$\textcolor{blue}{\bf{\checkmark}}$" indicating 
that TMD-factorization is valid and "$\textcolor{red}{\bf{!!}}$" indicating that it is problematic.
\begin{center}
\renewcommand{\labelitemi}{$$\textcolor{blue}{\bf{\checkmark}}$$}
\begin{itemize}
\item Back-to-back hadron or jet production in $e^+ e^-$-annihilation.
\end{itemize}
\renewcommand{\labelitemi}{$\textcolor{blue}{\bf{\checkmark}}$}
\begin{itemize}
\item Drell-Yan scattering.
\end{itemize}
\renewcommand{\labelitemi}{$\textcolor{blue}{\bf{\checkmark}}$}
\begin{itemize}
\item Semi-inclusive deep inelastic scattering ($e^+ + p \to H_1 + X$).
\end{itemize}
\renewcommand{\labelitemi}{$$\textcolor{red}{\bf{!!}}$$}
\begin{itemize}
\item \sout{Hadro-production of back-to-back jets or hadrons  ($H_1 + H_2 \to H_3 + H_4 + X$)}.
\end{itemize}
\end{center}

\section{Consistent Definitions and Evolution}

Above we emphasized the importance of precise TMD definitions  
for establishing the existence of TMD-factorization theorems.  
A consistent and agreed-upon set of definitions for the TMDs 
will also be a critical aspect of good phenomenology in the future because they are
needed for a correct treatment of evolution.
 
One may look to the more familiar collinear factorization formalism and its application 
to phenomenology for guidance on what is ultimately desired for a satisfactory TMD-factorization formalism.
In the collinear case, there is a clear conceptual path from naive parton 
model intuition and the full QCD treatment.  In inclusive DIS, for example, the parton model provides a basic 
factorization formula involving a (zeroth order) hard part and a collinear PDF.  The PDF is scale-independent.  
In transitioning to real QCD, the basic factorization structure remains 
valid, but the PDFs acquire scale dependence described by evolution equations.  Also, the full QCD factorization treatment provides the prescription for calculating higher orders in the hard part.  
Sophisticated fits have been performed for the collinear PDFs and FFs, and they are now essential tools
for good QCD phenomenology. 
 
To follow the same logical structure in treating TMD processes, we must first write down the 
parton model formulas.
For example, partonic reasoning gives the following $q_T$-dependent hadronic tensor for the DY cross section:
\begin{equation}
\label{eq:GPM1}
W^{\mu \nu} = \sum_f \, | H^{(0)}_f(Q) |^{\mu \nu}  \int d^2 {\bf k}_{1T} d^2 {\bf k}_{2T} \, F_{f/P_1}(x_1,k_{1T}) 
\bar{F}_{\bar{f}/P_2} (x_2, k_{2T}) \, \delta ({\bf k}_{1T} + {\bf k}_{2T} - {\bf q}_{T}).
\end{equation}
To complete the analogy with collinear factorization, what is needed is a set of definitions for the TMDs, $F_{f/P_1}(x_1,k_{1T})$ and $\bar{F}_{\bar{f}/P_2} (x_2, k_{2T})$ that allow Eq.~\eqref{eq:GPM1} to be rewritten in a way that
includes QCD effects, including the evolution of the TMDs, with minimal modification to the basic parton-model-like structure.
The definitions presented recently in Ref.~\cite{collins}, and which follow from the factorization derivation\footnote{To preserve 
space, we refer the reader directly to Ref.~\cite{collins} (chapter 11) for the explicit formulas for the definitions
and their explanation in terms of factorization.}, allow for this.  
The hadronic tensor is expressed in Ref.~\cite{collins} as
\begin{multline}
\label{eq:QCD}
W^{\mu \nu} = \sum_f \, | H_f(Q;\mu) |^{\mu \nu}  \\ \times \int d^2 {\bf k}_{1T} d^2 {\bf k}_{2T} \, F_{f/P_1}(x_1,k_{1T};\mu;\zeta_1) 
\bar{F}_{\bar{f}/P_2} (x_2, k_{2T};\mu;\zeta_2) \, \delta ({\bf k}_{1T} + {\bf k}_{2T} - {\bf q}_{T}) \\ + Y(Q,q_T) + \mathcal{O}(\Lambda/Q).
\end{multline}
Here, the hard part $H_f(Q;\mu)$ is calculable to arbitrary order and depends on the renormalization scale $\mu$.  
The TMDs also have scale dependence through $\mu$ and $\zeta_1, \zeta_2$.  The scales $\zeta_1$ and $\zeta_2$ are
related to the regulation of light-cone divergences and obey $\zeta_1 \zeta_2 = Q^4$.  The term $Y(Q,q_T)$ describes the matching to large $q_T$ where the approximations of TMD-factorization 
break down.
The first term on the right-hand side of Eq.~\eqref{eq:QCD} has the same general structure as Eq.~\eqref{eq:GPM1}.
The main differences are that: a.) the TMDs have acquired scale-dependence and, b.) the hard scattering can be calculated to arbitrary order in $\alpha_s$.  Note also that an explicit soft factor does \emph{not} appear (the role of soft gluons is now contained in the definitions of the TMDs).

The evolution of the individual TMDs now follows 
steps very close to the original CSS-formalism~\cite{CS1,CSS1}.  What remains is 
to systematically fit and tabulate them.  
Many of the ingredients for obtaining fits
already exist in the form of fixed scale fits, and in applications of the CSS formalism to the Drell-Yan 
cross section.  In Ref.~\cite{Aybat:2011zv}, a set of evolved unpolarized TMDs were constructed from fixed scale Gaussian fits to SIDIS data~\cite{Schweitzer:2010tt} and from fits to the Drell-Yan cross section 
in Ref.~\cite{Landry:2002ix,Konychev:2005iy}.  The evolution equations have thus allowed existing fits to be combined in a single unified formalism that describes the full range of $Q$.  

Having well-defined TMDs is also needed for an unambiguous method for calculating higher 
orders in the hard part.  
If we write the TMD-factorization formula schematically for DY as
\begin{equation}
\label{eq:fact}
W^{\mu \nu} = |H^2|^{\mu \nu} F_{f/P_1} \otimes \bar{F}_{\bar{f}/P_2},
\end{equation}
then the hard part is found, to arbitrary order, by calculating
\begin{equation}
\label{eq:fact2}
|H^2|^{\mu \nu} = \frac{W^{\mu \nu}}{F_{f/P_1} \otimes \bar{F}_{\bar{f}/P_2}}.
\end{equation}
In a perturbative expansion, the denominator in Eq.~\eqref{eq:fact2} becomes 
a series of double counting subtractions which ensure that $|H^2|^{\mu \nu}$ is perturbatively well-behaved.
With the definitions provided in Ref.~\cite{collins}, the calculation of $|H^2|^{\mu \nu}$ for 
SIDIS and DY becomes straightforward.  (See section 10.12.3.)
To order-$\alpha_s$ in the $\overline{\rm MS}$ scheme, they are:
\begin{eqnarray}
\left| \mathcal{H}_f(Q;\mu)^2 \right|_{\rm SIDIS} = 
e_f^2 \left| H_0^2 \right| \left(1 + \frac{C_F \alpha_s}{\pi} 
\left[\frac{3}{2} \ln \left( \frac{Q^2}{\mu^2} \right) - \frac{1}{2} \ln^2 \left( \frac{Q^2}{\mu^2} \right) - 4 \right] \right), \\
\left| \mathcal{H}_f(Q;\mu)^2 \right|_{\rm DY} = 
e_f^2 \left| H_0^2 \right| \left(1 + \frac{C_F \alpha_s}{\pi} 
\left[\frac{3}{2} \ln \left(\frac{Q^2}{\mu^2} \right) - \frac{1}{2} \ln^2 \left( \frac{Q^2}{\mu^2} \right) - 4 + 
\frac{\pi^2}{2} \right] \right).
\end{eqnarray}

\section{Future Directions}

We are now 
in a position to push the phenomenology further by extending the results to the case of spin-dependent 
functions like the Sivers and Boer-Mulders functions.  It will also be important to construct 
entirely new fits for the unpolarized case 
within the TMD-factorization treatment described above.  This will require calculating the various perturbatively calculable 
components to higher orders in $\alpha_s$, as well as calculating the $Y(Q,q_T)$-terms.   
These are all efforts we intend to continue.  Results and updates 
will be available at \url{http://projects.hepforge.org/tmd/}.

\begin{theacknowledgments}
We acknowledge J.~C.~Collins and P.~J.~Mulders for useful discussions.
Support was provided by the research 
program of the ``Stichting voor Fundamenteel Onderzoek der Materie (FOM)'', 
which is financially supported by the ``Nederlandse Organisatie voor Wetenschappelijk Onderzoek (NWO)''.
M.~Aybat also acknowledges support from the FP7 EU-programme HadronPhysics2 (contract no 2866403).
\end{theacknowledgments}

\bibliographystyle{aipproc}  

\bibliography{disbib}

\end{document}